\documentstyle[twoside,fleqn,espcrc2,epsf]{article}

\newcommand{\AmS}{{\protect\the\textfont2
  A\kern-.1667em\lower.5ex\hbox{M}\kern-.125emS}}

\title{  Chirality tubes along monopole trajectories 
         \thanks{Supported in part by FWF under Contract No. P11456} 
}

\author{Harald Markum, Wolfgang Sakuler and Stefan Thurner \\
\vspace{3mm}
Institut f\"{u}r Kernphysik, TU Wien,
         Wiedner Hauptstra\ss e 8-10, A-1040 Vienna, Austria\\ 
}

\begin{document}

\begin{abstract}
We classify the lattice by elementary 3-cubes which are associated  to 
dual links occupied by, or free of monopoles. We then compute the quark 
condensate, the quark charge and the chiral density on those cubes. 
By looking at distributions we demonstrate that monopole trajectories carry 
considerably more chirality with respect to the free vacuum. 
\end{abstract}

\maketitle
During the last years one has gained some insight into the mutual
interrelations of two distinct excitations of the QCD vacuum: 
monopoles and instantons. Both of those objects have been 
used to explain a wide variety of basic QCD properties, such as 
quark confinement, chiral symmetry breaking \cite{SHU88}  and the  $U_A(1)$ problem. 
The first property is usually associated with monopoles,  
the later ones with instantons. 
Instantons have integer topological charge $Q$ which is related to  
the chiral zero eigenvalues of the fermionic matrix 
with  a gauge field configuration 
via the Atiyah-Singer index theorem.
Since instantons carry chirality, and on the other hand 
it has been demonstrated that instantons are predominantly 
localized at regions where monopoles exist, the question arises 
whether monopoles carry chirality themselves. 
For calorons it has been proven  that they consist of 
monopoles \cite{baal} which might be a sign that  monopoles 
are indeed carriers of chirality. 

In this contribution we discuss this issue by directly looking at the 
chirality located on  monopole loops, and comparing it to 
the background. 
We do this by measuring conditional probability distributions 
of fermionic observables  of the form
$ \bar\psi\Gamma\psi$ with $\Gamma = 1,\gamma_4,\gamma_5$
in a standard staggered fermion setting.
Those quantities are usually referred to as the  
quark condensate, quark charge density, and the chiral density. 
Mathematically and numerically  
the local quark condensate $\bar \psi \psi (x)$ 
is a diagonal element of the inverse of the fermionic matrix
of the QCD action. The other fermionic operators 
are obtained by inserting the Euclidian $\gamma_4$ 
and $\gamma_5$ matrices. 

Under a conditional probability distribution we understand the  
probability of encountering a certain value for a fermionic observable  
$ \bar\psi\Gamma\psi$, under the condition that the local position is 
close to (or away from) a monopole trajectory,   
\begin{equation}
P^{\rm (no)mon}_{ s/t}(\bar\psi\Gamma\psi,x ) |_{x \in (\not\in) {\rm monopole \, tube}} \, ,
\end{equation}
where 
$x$ is indicating the local position, and $s/t$  
space- or time-like monopole trajectories.  
The core of the  monopole tube is the singular monopole trajectory, 
living on dual links, as obtained by the standard definition of monopoles in 
SU(3). We did not distinguish between the two independent 
colors of monopoles. 
For each dual link occupied by a monopole trajectory 
there exists an elementary 3-cube. 
The 8 sites of such a cube constitute the section of the monopole tube 
corresponding to that dual link. 

Our simulations were performed for full SU(3) QCD on an 
$8^{3} \times 4$ lattice with
periodic boundary conditions.  
Dynamical quarks in Kogut-Susskind discretization   
with $n_f=3$ flavors of degenerate  mass $m=0.1$ were taken into account using the 
pseudofermionic method. 
We performed runs  in the confinement phase at $\beta=5.2$. 
Measurements were taken on 2000 configurations separated 
by 50 sweeps.

We computed correlation functions between two observables 
${\cal O}_1(x)$ and ${\cal O}_2(y)$ \cite{boulder}, 
\begin{equation}
\label{correlations}
g(y-x)=\langle {\cal O}_1(x) {\cal O}_2(y) \rangle - 
       \langle {\cal O}_1\rangle \langle {\cal O}_2\rangle \, .
\end{equation}
In Fig. 1 we display results for 
${\cal O}_1$ a local ferm\-ionic observable (except in (d)) 
and $ {\cal O}_2$ the monopole charge density $\rho$.  
All correlations exhibit an extension  
of several lattice spacings and show an exponential 
falloff over the whole range.  
The corresponding screening masses are given 
in Table 1 in GeV for 3 levels of cooling. 
They are a 
coarse measure for the chirality profile of the monopole tube. 
It is apparent that cooling does not change the 
screening masses drastically. 

\begin{figure}
\epsfxsize=7.3cm\epsffile{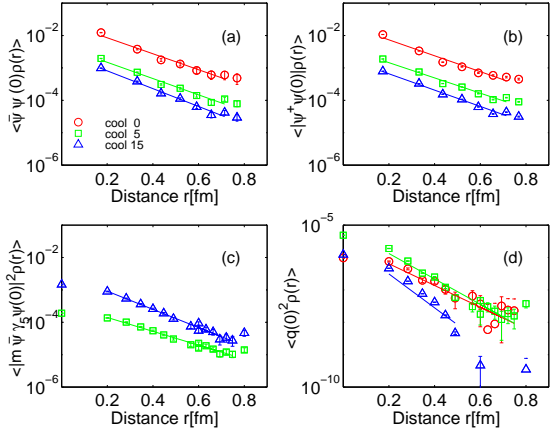} 
\vspace{-4mm}
\caption{ Correlation functions with the monopole density $\rho(r)$.  }
\label{corr}

\vspace{-4mm}
\end{figure}

For reasons of comparison we included 
in the table the screening masses for the correlation 
with the topological charge density (squared).
We find that the correlations of the color charge density 
$\bar\psi\psi(x)$ and
$|\psi^{\dagger}\psi(x)|$ with the topological
charge density are very similar,
both in the slopes and the absolute values. This becomes clear because the
quark condensate can be interpreted as the absolute value of the quark density.
However, cooling (or some other kind of smoothing)
is inevitable to obtain nontrivial correlations between the chiral density,
${\cal O}_1=\bar \psi \gamma_5 \psi(x)$, and the topological charge density.
This can be expected since both quantities are connected via the anomaly.
The topological charge of a gauge field is related to the
chiral density of the associated fermion field by
$Q=\int q(x) d^4x= m\int \bar\psi\gamma_5\psi(x)  d^4x$. 
We have checked that this relation also holds approximately for
the corresponding lattice observables on individual configurations.
The autocorrelation function of the density of the
topological charge $<q(0)q(r)>$ should be compared to
$<\bar\psi\gamma_5\psi(0)q(r)>$.
\begin{table}[t]
\caption{Screening masses in GeV from fits to exponential decays of
        the various correlators for several cooling steps.  }
\begin{tabular}{l c c c }
\hline
Correlation                          & cool 0   & cool 5   & cool 15  \\
\hline
$|\psi^{\dagger}\psi(0)| \rho(r)$    & 1.13(02) & 1.08(01) & 1.16(01) \\
$\bar\psi\psi(0) \rho(r)$            & 1.14(10) & 1.16(05) & 1.27(06) \\
$|\bar\psi\gamma_5\psi(0)|^2 \rho(r)$&  -       & 0.98(07) & 1.30(10) \\
$q^2(0) \rho(r)$                     & 1.54(47) & 1.81(20) & 2.41(58) \\
\hline
$|\psi^{\dagger}\psi(0)| q^2(r)$      & 1.25(66) & 1.29(10) & 1.42(09) \\
$\bar\psi\psi(0) q^2(r)$              & 1.32(34) & 1.38(16) & 1.47(16) \\
$\bar\psi \gamma_5\psi(0) q(r)$       &  -       & 0.84(02) & 0.48(01) \\
$q(0) q(r)$                           &  -       & 1.67(02) & 0.84(01) \\
\hline
\end{tabular}
\end{table}
If the classical t'Hooft in\-st\-an\-ton with size $\rho_I$
is considered,  the topological charge density is
\begin{equation}
q(x) \propto \rho_I^4(x^2+\rho_I^2)^{-4} \quad .
\end{equation}
On the other hand the corresponding density of fermionic quantities \cite{SHU88}
\begin{equation}
\bar \psi \psi(x)  \propto \bar \psi\gamma_5\psi(x)  \propto \rho_I^2(x^2+\rho_I^2)^{-3}
\end{equation}
is broader.
This behavior is reflected in Table 1 and means that the 
local relation $q(x)=m\bar\psi\gamma_5\psi(x)$ does not hold. 

\begin{figure}[t]
\begin{center}
\begin{tabular}{c}
\hspace{-7mm} \epsfxsize=7.8cm\epsffile{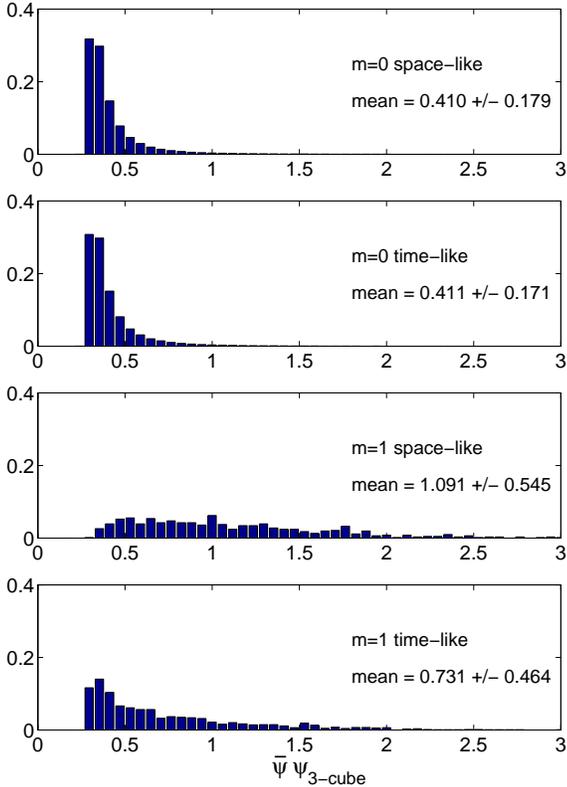} 
\end{tabular}
\end{center}

\vspace{-11mm}
\caption{
Conditional probability distributions for $\bar \psi \psi(x)$ 
	due to monopole appearance. 
}
\label{dist1}

\end{figure}
Figure~\ref{dist1} shows results for the 
conditional probability distributions for 
$\bar \psi \psi$ (in multiples of the quark mass) 
in the case of monopole presence (m=1)
or absence (m=0). The m=0 case exhibits a relatively narrow 
distribution of the fermionic quantity around $0.41$ with 
a variance of 0.179 or 0.171, for space- and time-like 
monopole trajectories respectively. 
The m=1 case is clearly different and shows a much broader 
distribution, with both the mean and the variance being 
about a factor two larger. The time-like trajectories  
yield distributions which are still peaked on the left, 
like in the m=0 case, whereas this is not observed for 
space-like trajectories.  
 
\begin{figure}[t]
\begin{center}
\hspace{-5mm} \epsfxsize=7.8cm\epsffile{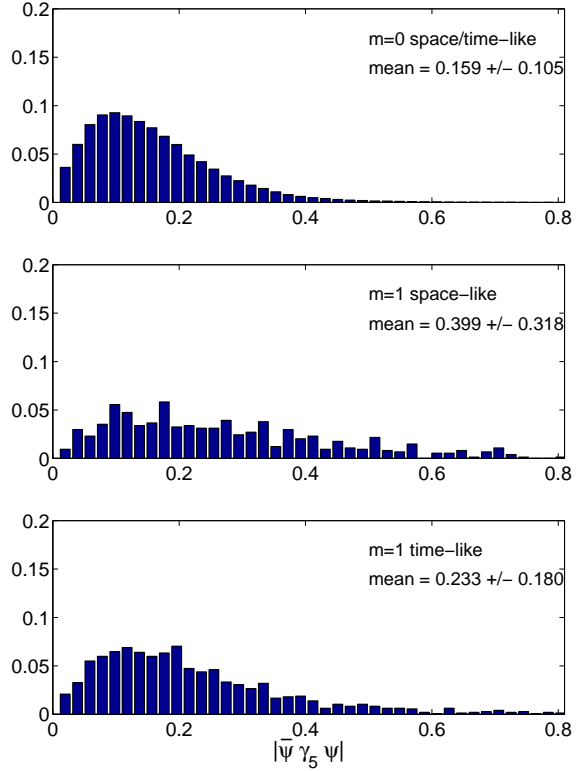} 
\end{center}

\vspace{-10mm}
\caption{
Conditional probability distributions for 
$|\bar \psi \gamma_5 \psi(x)|$ due to monopole appearance. 
}
\vspace{-2mm}
\label{dist2}
\end{figure}

The plot suggests 
that in the close neighborhood  of a monopole it becomes more likely 
to encounter large values of $\bar \psi \psi(x)$.
The form of the distributions  
points towards a picture where monopole tubes  
carry a space-time dependent  density of the fermionic observables. 
In the confinement, space-like tubes bear more chirality. 
The same situation is found for the other observables 
${\rm Re}(\psi^{\dagger} \psi)$ and 
$|\bar \psi \gamma_5 \psi|$ (see Fig.~\ref{dist2}).  
The figures depict the situation after 
15 cooling steps. We checked that these observations 
can be made also after 5 cooling steps. 

In summary, 
the computations of correlation functions
between the monopole charge  density and the fermionic observables 
yield an exponential decrease. The 
screening masses correspond to those of correlators between the topological 
charge density and the same fermionic observables \cite{boulder}. 
Our calculations of conditional distribution functions 
of fermionic observables point out a significant enhancement for  
finding large chirality  in the neighborhood of monopole trajectories. 
The same distributions also indicate that the monopoles are not 
covered by a uniform chirality tube. 

%


\begin{thebibliography}{99}
%
\bibitem{SHU88}
E.V.~Shuryak,
Nucl.~Phys.~{ B302} (1988) 559;
T.~Sch\"{a}fer and E.V.~Shuryak, 
Rev.~Mod.~Phys. 70 (1998) 323. 
%
\bibitem{baal}
T.C. Kraan and P. van Baal,  
Nucl.~Phys.~{B} (Proc. Suppl.) 73 (1999) 554.
%
\bibitem{boulder} 
H. Markum, W. Sakuler and S. Thurner, 
Nucl.~Phys. {B} (Proc. Suppl.) 73 (1999) 509.
%

\end{thebibliography}
\end{document}